\RequirePackage{ifpdf}
\pdfoutput=1
\documentclass[11pt,a4paper]{article}
\usepackage{jheppub}
\usepackage{graphicx}
\usepackage{amssymb}
\usepackage{amsmath}
\usepackage{amsthm}
\usepackage{slashed}
\usepackage{color,rotating}
\usepackage{bbm}
\usepackage{array}
\usepackage{colonequals}
\usepackage[utf8]{inputenc}
\usepackage{subcaption}
\usepackage{xcolor}
\usepackage{stackrel}
\DeclareMathOperator{\tr}{tr}
\allowdisplaybreaks

\usepackage[normalem]{ulem}

\newcommand{\be}{\begin{equation}}
\newcommand{\ee}{\end{equation}}

\newcommand{\UV}{{\small UV}}
\newcommand{\IR}{{\small IR}}
\newcommand{\QFT}{{\small QFT}}

\newcommand{\IDG}{{\small IDG}}
\newcommand{\GR}{{\small GR}}
\newcommand{\RG}{{\small RG}}
\newcommand{\FRG}{{\small FRG}}
\newcommand{\LHC}{{\small LHC}}
\newcommand{\QED}{{\small QED}}

\newcommand{\eg}{{\textit{e.g.}}}
\newcommand{\ie}{{\textit{i.e.}}}

\newcommand{\mans}{\ensuremath{\mathfrak{s}}}
\newcommand{\mant}{\ensuremath{\mathfrak{t}}}
\newcommand{\manu}{\ensuremath{\mathfrak{u}}}

\newcommand{\cA}{\ensuremath{\mathcal{A}}}

\title{Graviton-Mediated Scattering Amplitudes \\ from the Quantum Effective Action}
\author[a]{Tom Draper\,\href{https://orcid.org/0000-0002-6895-894X}{\protect \includegraphics[scale=.07]{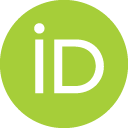}},}
\author[b]{Benjamin Knorr\,\href{https://orcid.org/0000-0001-6700-6501}{\protect \includegraphics[scale=.07]{ORCIDiD_icon128x128.png}},}
\author[c]{Chris Ripken\,\href{https://orcid.org/0000-0003-2545-5047}{\protect \includegraphics[scale=.07]{ORCIDiD_icon128x128.png}}}
\author[a]{and Frank Saueressig\,\href{https://orcid.org/0000-0002-2492-8271}{\protect \includegraphics[scale=.07]{ORCIDiD_icon128x128.png}}}

\affiliation[a]{Institute for Mathematics, Astrophysics and Particle Physics (IMAPP), \\
Radboud University Nijmegen, Heyendaalseweg 135, 6525 AJ Nijmegen, The Netherlands}
\affiliation[b]{Perimeter Institute for Theoretical Physics, 31 Caroline St. N., Waterloo, ON N2L 2Y5, Canada}
\affiliation[c]{Institute of Physics (THEP), University of Mainz, Staudingerweg 7, 55128 Mainz, Germany}

\emailAdd{t.draper@student.ru.nl}
\emailAdd{bknorr@perimeterinstitute.ca}
\emailAdd{aripken@uni-mainz.de}
\emailAdd{f.saueressig@hef.ru.nl}

\abstract{We employ the curvature expansion of the quantum effective action for gravity-matter systems
to construct graviton-mediated scattering amplitudes for non-minimally coupled scalar fields in a  Minkowski background. By design, the formalism parameterises all quantum corrections to these processes and is manifestly gauge-invariant.
	The conditions resulting from \UV{}-finiteness, unitarity, and causality are analysed in detail and it is shown by explicit construction that the quantum effective action provides sufficient room to meet these structural requirements without introducing non-localities or higher-spin degrees of freedom. Our framework provides a bottom-up approach to all quantum gravity programs seeking for the quantisation of gravity within the framework of quantum field theory. Its scope is illustrated by specific examples, including effective field theory, Stelle gravity, infinite derivative gravity, and Asymptotic Safety.}


\begin{document}
\maketitle
\section{Introduction}\label{sect:intro}

Quantum field theories have been extremely successful in providing predictions in high-energy physics. They constitute the framework for formulating the Standard Model of particle physics, which has been tested with great precision. In particular, with the discovery of the Higgs boson, experiments at the \LHC{} have confirmed all particles in the Standard Model. In order to relate experiment to theory, scattering amplitudes play a key role. On the one hand, they form the basis for computing experimentally accessible quantities such as differential cross sections and decay rates. On the other hand, theoretical requirements such as unitarity, causality and positivity put constraints on the amplitudes \cite{Froissart:1961ux, Froissart:2010, Cerulus:1964cjb,Epstein:2019zdn, Adams:2006sv, Camanho:2014apa, Bellazzini:2015cra, Chandrasekaran:2018qmx}.\footnote{For related ideas in the context of string theory see \cite{Vafa:2005ui,Ooguri:2006in} as well as the recent review \cite{Palti:2019pca}.}

At this stage, it is an open debate under which conditions gravity may be formulated as a quantum field theory. Applying the quantisation techniques successful for the standard model of particle physics to general relativity (\GR{}) yields a perturbatively non-renormalisable theory \cite{'tHooft:1974bx, Goroff:1985sz, Goroff:1985th}. This is reflected in the amplitude of gravity-mediated scattering of matter; for a two-to-two particle process, the tree-level amplitude diverges quadratically with the centre-of-mass energy. This divergence is aggravated by adding loop corrections \cite{Polchinski:1998rq}. Therefore, it is commonly accepted that \GR{} is an effective field theory valid up to the Planck scale $M_{\text{Pl}} \simeq 10^{19} \text{ GeV}$ \cite{Donoghue:1993eb,Donoghue:1994dn,Burgess:2003jk,Donoghue:2017pgk}. By now, there is a significant number of proposals on how to complete \GR{} into a fundamental theory valid on all scales with String Theory \cite{Becker:2007zj, Schomerus:2017lqg}, Loop Quantum Gravity \cite{Thiemann:2002nj, Rovelli:2004tv,Ashtekar:2017yom}, Asymptotic Safety \cite{Niedermaier:2006wt, Litim:2011cp, Reuter:2012id, Ashtekar:2014kba, Eichhorn:2017egq, Eichhorn:2018yfc, Reichert:2020mja}, Causal Dynamical Triangulations \cite{Ambjorn:2012jv,Loll:2019rdj}, Causal Set Theory \cite{Sorkin:2003bx, Surya:2011yh}, Group Field Theory \cite{Baratin:2010wi, Oriti:2011jm}, infinite derivative gravity \cite{Biswas:2011ar,Talaganis:2016ovm, Talaganis:2014ida, Buoninfante:2018mre}, and weakly non-local gravity \cite{Modesto:2017sdr} constituting only some selected examples.

In order to systematically study scattering amplitudes in a way that is agnostic to the underlying microscopic physics, we build on a parameterisation of the quantum effective action $\Gamma$. The scattering amplitudes are then calculated from the
tree-level Feynman diagrams constructed from the propagators and vertices encoded in $\Gamma$. Although the computation of $\Gamma$ is generically difficult, we can study parameterisations capturing all contributions to a physical process. In particular, the high-energy behaviour of the scattering amplitudes can  be analysed based on a curvature expansion of 
 $\Gamma$ including form factors \cite{Barvinsky:1990up} (also see \cite{Knorr:2019atm} for a detailed discussion). The latter can be regarded as the operator equivalent of momentum-dependent interactions which generalise Minkowski-space interactions to curved spacetime. The main result of this paper is the most general (manifestly gauge-invariant) scattering amplitude for a gravity-mediated two-to-two scalar particle scattering in a Minkowski background.
On this basis, we derive conditions for the gravitational and matter form factors in order to obey unitarity, causality, and positivity. This enables us to identify classes of quantum effective actions compatible with these fundamental properties. 

Taking such a general approach allows us to study different quantum gravity theories  by specifying the corresponding form factors.
In this way, effective field theory, Stelle gravity, infinite-derivative gravity, Asymptotic Safety, and renormalisation group improvements can be treated in a uniform language.
By considering their pole structure and high-energy behaviour, we investigate their compatibility with unitarity and causality. This leads to a specific model where the amplitudes become scale-free and compatible with all fundamental requirements 
without introducing new particle resonances. This model is described in detail in \cite{Draper:2020bop}.

The rest of this paper is structured as follows. In \autoref{sect:eff_action}, we present the effective action that generates the scattering amplitudes. In \autoref{sect:gravityscalarscalarscattering}, we compute the most general four-scalar scattering amplitude and the resulting differential cross section. The key properties of our result are analysed in \autoref{sec:generalscattering}. These sections contain the main novel results of this paper. In \autoref{sect:physics_of_s_channel}, we specialise our result to various quantum gravity models studied in the literature giving their description in terms of form factors. We conclude the paper with a summary and a discussion in \autoref{sect:summary}. The explicit form of the propagators and  vertices entering the computation is given in \autoref{App.A} and we summarise our conventions in \autoref{sect:conventions}. Throughout this work we study the scattering of scalar matter only. The generalisation to other matter fields is left for future work.

\section{ The quantum effective action including form factors}\label{sect:eff_action}

We start our investigation with a brief introduction to form factors in curved spacetime. A detailed discussion can be found in \cite{Knorr:2019atm}.
Form factors arise naturally in the computation of loop corrections. A textbook example is the electron self-energy in \QED{} which introduces a non-trivial momentum dependence in the electron propagator. Generically, form factors are functions of momentum invariants entering  the propagators and interaction vertices.
In flat spacetime, the Fourier transform allows to switch from the momentum-space representation of the form factors to a position-space representation where the corresponding functions  depend on partial derivatives. The correspondence principle then provides a natural generalisation to curved space, replacing the partial derivatives by covariant derivatives.

\subsection{Construction of the generic gravity-scalar action}
We will now use form factors to construct a systematic curvature expansion of the quantum effective action
\be\label{eq:Gans1}
\Gamma = \Gamma_\text{grav} + \Gamma_\text{gf} + \Gamma_\text{matter} \, . 
\ee
This will provide the basis for computing the amplitude for two-to-two scalar scattering processes in a flat background.

\paragraph{Gravitational action}
We begin with the purely gravitational part. In this case, an efficient expansion scheme is in powers of the curvature tensor $\mathcal{R}$. This expansion  is expected to be accurate if $D^2 \mathcal{R} \gg \mathcal{R}^2$, \eg{}, for near-flat spacetime. Up to second order in the curvature, the purely gravitational part reads
\begin{equation}
	\label{eq:Gammagrav}
 \Gamma_\text{grav} = \frac{1}{16\pi G_N} \int \text{d}^4x \, \sqrt{-g} \left[ -R - \frac{1}{6} R \, f_{RR}(\Delta) \, R + \frac{1}{2} C_{\mu\nu\rho\sigma} \, f_{CC}(\Delta) \, C^{\mu\nu\rho\sigma} + \mathcal O(\mathcal{R}^3) \right] \, .
\end{equation}
Here $G_N$ is Newton's constant, $\Delta = - g^{\mu\nu} D_\mu D_\nu$ is the d'Alembertian and $f_{RR}$ and $f_{CC}$ are the form factors which determine the flat-spacetime graviton propagator. The normalisation is chosen for convenience, and signs are such that if the form factors are positive, no additional poles appear in the propagator. In the remainder of this work, all dimensionful quantities are measured in units of the Planck mass $M_{\rm Pl} \equiv G_N^{-1/2}$, which corresponds to setting $G_N = 1$.

A few remarks are in order here. For simplicity, we have set the cosmological constant to zero. This ensures that flat Minkowski spacetime is an on-shell solution to the equations of motion. While the Minkowski background is motivated for technical (momentum-space techniques) as well as conceptual reasons (flat spacetime is a good approximation for earth-based physics) it is \emph{not} necessary that flat spacetime is a solution to the equations of motion. Since we discuss scalar scattering only, the only gravitons that appear are virtual and can thus remain off-shell.

In principle, one could also include the Gauss-Bonnet term. However, in 3+1 spacetime dimensions this reduces to a topological surface term and does not contribute to the dynamics. Furthermore, one could think of adding a form factor associated to the tensor structure with two Ricci tensors. Using the Bianchi identity this can be mapped onto the tensor structures in \eqref{eq:Gammagrav} and is therefore not present \cite{Codello:2012kq,Knorr:2019atm}.

In order to obtain a well-defined graviton propagator, we include the gauge-fixing action
\begin{equation}
	\label{eq:gaugefixing}
 \Gamma_\text{gf} = \frac{1}{32\pi \alpha} \int \text{d}^4x \, \sqrt{-\eta} \left( \partial^\mu h_{\mu\nu} - \frac{1+\beta}{4} \partial_\nu h \right) \left( \partial_\rho h^{\rho\nu} - \frac{1+\beta}{4} \partial^\nu h \right) \, .
\end{equation}
Here $h_{\mu\nu}$ is the metric fluctuation around the Minkowski metric $\eta$,
\begin{equation}
 g_{\mu\nu} = \eta_{\mu\nu} + h_{\mu\nu} \, .
\end{equation}
We will leave the gauge parameters $\alpha$ and $\beta$ general to see the gauge independence of the scattering amplitude explicitly. It is straightforward to calculate the flat propagator in a general gauge \cite{Knorr:2017fus}. For the convenience of the reader, the result is reproduced in Appendix \ref{app:gravitonpropagator}. In principle, gauge fixing also generates an action for the Faddeev-Popov ghosts. These do not contribute to the scalar scattering
and are therefore irrelevant in the present construction.

\paragraph{Scalar action}
We now construct the gravity-matter sector of the action for two  distinct scalar fields $\phi$ and $\chi$. We assume that the scalars each possess a $\mathbbm Z_2$-symmetry, and that they are both in the symmetric phase. The matter sector is then given by
\be\label{eq:gamma-matter}
\Gamma_{\rm matter} = \Gamma_\phi + \Gamma_{\phi^4} + \Gamma_\chi + \Gamma_{\chi^4} + \Gamma_{\phi^2\chi^2} \, , 
\ee
with the building blocks $\Gamma_\phi$, $\Gamma_{\phi^4}$, and $\Gamma_{\phi^2\chi^2}$ defined in eqs.\ \eqref{eq:gamma_scal}, \eqref{eq:gammaff}, and \eqref{eq:gammafc}, respectively. The actions for $\chi$ are obtained by replacing $\phi \to \chi$. Notably, eq. \eqref{eq:gamma-matter} is one-loop complete, in the sense that it contains all interaction monomials that appear in the one-loop effective action \cite{'tHooft:1974bx,Donoghue:1994dn} and contribute to the scattering processes analysed in this work.\footnote{Recasting the matter interactions found in the one-loop effective field theory \cite{Satz:2010uu,Ohta:2020bsc} into the form-factor parameterisation underlying this work gives rise to non-minimal couplings of order $\mathcal R^2$ and higher arising from commutators of covariant derivatives. However, these terms do not contribute to the scattering amplitude in a flat Minkowski background.}

For a single scalar field, the action contributing to a two-to-two particle scattering mediated by a graviton reads
\begin{equation}\label{eq:gamma_scal}
\begin{aligned}
 \Gamma_\phi &= \int \text{d}^4x \sqrt{-g} \Bigg[ \frac{1}{2} \phi \, f_{\phi\phi}(\Delta) \, \phi + f_{R\phi\phi}(\Delta_1,\Delta_2,\Delta_3) \, R \, \phi \, \phi \\
 &\qquad\qquad\qquad + f_{Ric\phi\phi}(\Delta_1,\Delta_2,\Delta_3) \, R^{\mu\nu} \, (D_\mu D_\nu \phi) \, \phi 
 + \mathcal O(\mathcal{R}^2,\phi^4) \Bigg] 
 \, .
\end{aligned}
\end{equation}
For the multi-argument form factors, the index on the d'Alembertian indicates the object on which it acts, \eg{},
\begin{equation}
 f(\Delta_1,\Delta_2,\Delta_3) = \Delta_1^{n_1} \Delta_2^{n_2} \Delta_3^{n_3} \; \Rightarrow \; f(\Delta_1,\Delta_2,\Delta_3) X \, Y \, Z = \left( \Delta^{n_1} X \right) \left( \Delta^{n_2} Y \right) \left( \Delta^{n_3} Z \right) \, .
\end{equation}
The techniques needed for the derivation of the two-scalar-graviton vertex from this action have been discussed in detail in \cite{Knorr:2019atm}. In Appendix \ref{app:vertices} we write out the result.

The self-interactions consist of four-point vertices.
For distinct scalars, the action reads
\begin{equation}\label{eq:gammafc}
			\Gamma_{\phi^2\chi^2} = \frac{1}{4}	\int \text{d}^4x	\sqrt{-g}	f_{\phi^2\chi^2}\left(\{-D_i\cdot D_j\}_{i< j}\right)	\phi^2\chi^2
			\,,
\end{equation}
where the indices $i,j$ run from $1$ to $4$. The form factor $f_{\phi^2\chi^2}$ is symmetric upon interchanging indices $1 \leftrightarrow 2$ and $3 \leftrightarrow 4$. For identical particles, the four-point interaction has the form
\begin{equation}\label{eq:gammaff}
			\Gamma_{\phi^4} = \frac{1}{4!}	\int \text{d}^4x	\sqrt{-g}	f_{\phi^4}\left(\{-D_i\cdot D_j\}_{i< j}\right)	\phi^4
			\,,
\end{equation}
which is now by definition symmetric in all its arguments. Interactions of the form \eqref{eq:gamma_scal} with one of the $\phi$'s replaced by $\chi$ are forbidden by our assumption of $\mathbb{Z}_2$-symmetry.

Following standard procedure, we define the on-shell conditions on the asymptotic states by the free part of the equations of motion. The linearised equation of motion for $\phi$ on a flat background reads
\begin{equation}
 f_{\phi\phi}(-\partial^2) \phi = 0 \, .
\end{equation}
We will assume that $f_{\phi\phi}$ has a unique zero, defining the mass of the scalar field,
\begin{equation}
 f_{\phi\phi}(m_\phi^2) = 0 \, ,
\end{equation}
and we will canonically normalise the field so that it has a standard kinetic term on-shell, that is
\begin{equation}\label{eq:norm_cond_kinetic}
 f_{\phi\phi}'(m_\phi^2) = 1 \, .
\end{equation}
We assume that a similar equation of motion holds for $\chi$, but allow for distinct masses $m_\phi \neq m_\chi$.

\section{Gravity-mediated scalar-scalar scattering amplitude}\label{sect:gravityscalarscalarscattering}

We will now present the scattering amplitudes of two-to-two scalar scattering processes described by \eqref{eq:Gans1}. First, we will focus on distinguishable fields, scattering in the process $\phi\phi \to \chi\chi$. After that, we move on to scattering of identical particles.

\subsection{\texorpdfstring{$\phi\phi\to\chi\chi$}{distinct scalar-scalar} scattering}

\begin{figure}
 \centering
 \includegraphics[width=0.4\textwidth]{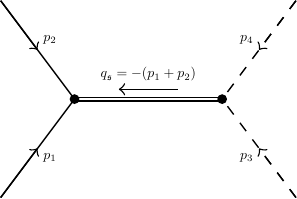}
 \caption{The \mans{}-channel Feynman diagram that contributes to the $\phi\phi\to\chi\chi$ scattering. The external solid lines correspond to $\phi$-legs, external dashed lines correspond to $\chi$-legs, the internal double line corresponds to the gauge-fixed graviton propagator obtained from \eqref{eq:Gans1}. The black dots indicate the three-point $h\phi\phi$- and $h\chi\chi$-vertices encoded in \eqref{eq:gamma_scal}.}
 \label{fig:diagram_two_fields}
\end{figure}

The amplitude associated with $\phi\phi\to \chi\chi$ scattering receives contributions from
graviton-mediated scattering, $\mathcal A_\mans^{\phi\chi}$, and matter self-interactions $\mathcal A_4^{\phi\chi}$,
\be	
\mathcal A^{\phi\chi} = \mathcal A_\mans^{\phi\chi} + \mathcal A_4^{\phi\chi}.
\ee
The gravity-mediated contribution is encoded in the Feynman diagram shown in \autoref{fig:diagram_two_fields}. Using the terminology associated with the Mandelstam variables $\mans$, $\mant$ and $\manu$ introduced in eq.\ \eqref{eq:mansdef}, this corresponds to $\mans$-channel scattering. 
The amplitude of this diagram is computed by combining the graviton propagator \eqref{eq:grav_propagator} and the vertices \eqref{eq:vertex}.
The calculation was performed with the help of the Mathematica package suite \emph{xAct} \cite{xActwebpage,2007CoPhC.177..640M,Brizuela:2008ra,2008CoPhC.179..597M,2014CoPhC.185.1719N}, yielding
\begin{equation}\label{eq:mixed_amp}
\begin{aligned}
 \mathcal A_\mans^{\phi\chi} &= \frac{4\pi}{3} \Bigg[ - \left(1+ \mans f_{Ric\phi\phi}(\mans,m_\phi^2,m_\phi^2) \right) \left(1+ \mans f_{Ric\chi\chi}(\mans,m_\chi^2,m_\chi^2) \right) G_{CC}(\mans) \\
  &\hspace{6.5cm} \times \left\{ \mant^2 - 4 \mant \manu + \manu^2 + 2\left( m_\phi^2 - m_\chi^2 \right)^2 \right\} \\
  &\hspace{0.5cm} + \left( (\mans + 2m_\phi^2) (1 + \mans f_{Ric\phi\phi}(\mans,m_\phi^2,m_\phi^2)) - 12 \mans f_{R\phi\phi}(\mans,m_\phi^2,m_\phi^2) \right) \\
  &\hspace{1.2cm} \times \left( (\mans + 2m_\chi^2) (1 + \mans f_{Ric\chi\chi}(\mans,m_\chi^2,m_\chi^2)) - 12 \mans f_{R\chi\chi}(\mans,m_\chi^2,m_\chi^2) \right) G_{RR}(\mans) \Bigg] \, .
\end{aligned}
\end{equation}
The Mandelstam variables are subject to the relation \eqref{eq:msrel}. Kinematically we have to assume that the energy transfer is at least the difference of squared final and squared initial mass, if the former is larger. Notably, the scalar kinetic form factors $f_{\phi\phi}$ and $f_{\chi\chi}$ do not contribute to $\mathcal A_\mans^{\phi\chi}$. This is because any instance of $f_{\phi\phi}$ or $f_{\chi\chi}$ is evaluated on-shell, and thus either vanishes if the form factor itself appears, or gives a factor of one by the normalisation condition \eqref{eq:norm_cond_kinetic}.
	Moreover, $\mathcal A^{\phi\chi}_\mans$ does not depend on the gauge fixing parameters $\alpha$ and $\beta$. The contraction of the gauge-fixed graviton propagator with the three-point vertices together with the on-shell conditions  projects the former on its gauge-invariant part. This establishes the gauge invariance of our result, which is an essential feature of any observable.
The amplitude related to the matter self-interactions reads
\begin{equation}\label{eq:4mixed}
			\mathcal A_4^{\phi\chi}
	=
			f_{\phi^2\chi^2}\left(
				\tfrac{\mans-2m_\phi^2}{2},
				\tfrac{\mant-m_\phi^2-m_\chi^2}{2},
				\tfrac{\manu-m_\phi^2-m_\chi^2}{2},
				\tfrac{\manu-m_\phi^2-m_\chi^2}{2},
				\tfrac{\mant-m_\phi^2-m_\chi^2}{2},
				\tfrac{\mans-2m_\chi^2}{2}
			\right) \,.
\end{equation}

 By crossing symmetry, the scattering amplitude for the process $\phi\chi \to \phi\chi$ is obtained by interchanging $\mans \leftrightarrow \mant$.

\subsection{\texorpdfstring{$\phi\phi\to\phi\phi$}{single scalar-scalar} scattering}

\begin{figure}[t]
 \centering
 \begin{subfigure}[b]{0.3\textwidth}
  \centering
  \includegraphics[width=\textwidth]{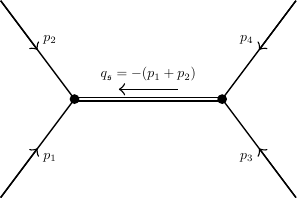}
  \caption{\mans{}-channel}
  \label{fig:diagrams_one_field_s}
 \end{subfigure}
 ~
 \begin{subfigure}[b]{0.3\textwidth}
  \centering
  \includegraphics[height=\textwidth]{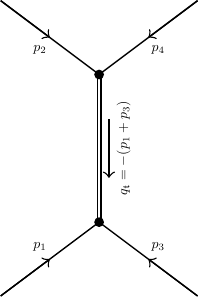}
  \caption{\mant{}-channel}
  \label{fig:diagrams_one_field_t}
 \end{subfigure}
 ~
 \begin{subfigure}[b]{0.3\textwidth}
  \centering
  \includegraphics[height=\textwidth]{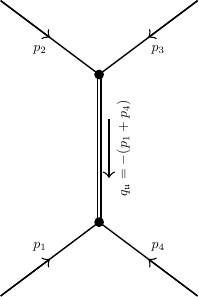}
  \caption{\manu{}-channel}
  \label{fig:diagrams_one_field_u}
 \end{subfigure}
 \caption{Feynman diagrams encoding the graviton-mediated contribution to the $\phi\phi\to\phi\phi$ scattering amplitude. The external lines correspond to $\phi$-legs, the internal double line corresponds to the 
 gauge-fixed graviton propagator obtained from \eqref{eq:Gans1}, and the black dots indicate the three-point $h\phi\phi$-vertex encoded in \eqref{eq:gamma_scal}.	
 	}
 \label{fig:diagrams_one_field}
\end{figure}

Let us briefly discuss the scattering amplitude for a single scalar field, $\phi\phi \rightarrow \phi\phi$. In this process, there are three different Feynman diagrams involving the exchange of a virtual graviton. These diagrams are shown in \autoref{fig:diagrams_one_field} and correspond to the $\mans$-, $\mant$- and $\manu$-channel. The amplitude of the $\mans$-channel diagram is straightforwardly obtained from \eqref{eq:mixed_amp} by making the identifications
\begin{equation}
		m_\chi\to m_\phi
		\,, \qquad
		f_{Ric\chi\chi}\to f_{Ric\phi\phi}
		\,, \qquad
		f_{R\chi\chi}\to f_{R\phi\phi}
		\,.
\end{equation}
The other diagrams are then obtained by crossing symmetry; the $\mant$- and $\manu$-channel diagrams are obtained from the $\mans$-channel diagram by interchanging $\mans \leftrightarrow \mant$ and $\mans \leftrightarrow \manu$, respectively. As an alternative check, we have performed the explicit computation of all channels. This results in
\be
\cA^{\phi\phi} = \mathcal A_\mans^{\phi\phi} + \mathcal A_\mant^{\phi\phi} + \mathcal A_\manu^{\phi\phi} + \mathcal A_4^{\phi\phi} \, ,
\ee
with the building blocks being
\begin{align}
 \mathcal A_\mans^{\phi\phi} &= \frac{4\pi}{3} \Bigg[ - \left(1+ \mans f_{Ric\phi\phi}(\mans,m_\phi^2,m_\phi^2) \right)^2 G_{CC}(\mans) \left\{ \mant^2 - 4 \mant \manu + \manu^2 \right\} \notag \\
 &\hspace{0.5cm} + \Big( (\mans+2m_\phi^2)\left( 1+\mans f_{Ric\phi\phi}(\mans,m_\phi^2,m_\phi^2)\right) - 12 \mans f_{R\phi\phi}(\mans,m_\phi^2,m_\phi^2) \Big)^2 G_{RR}(\mans) \Bigg] \, , \label{eq:Amp_phi_gen}
\end{align}
and
\begin{equation}\label{eq:phi4amplitude}
 \mathcal A_4^{\phi\phi} = f_{\phi^4}\left( \frac{\mans-2m_\phi^2}{2}, \frac{\mant-2m_\phi^2}{2}, \frac{\manu-2m_\phi^2}{2}, \frac{\manu-2m_\phi^2}{2}, \frac{\mant-2m_\phi^2}{2}, \frac{\mans-2m_\phi^2}{2} \right) \, .
\end{equation}
This completes our computation of the scattering amplitudes.

\section{The physics of graviton-mediated scattering -- general case}\label{sec:generalscattering}
In this section, we study the properties of the scattering amplitudes computed in \autoref{sect:gravityscalarscalarscattering}. In \autoref{sect:gen_s_channel}, we consider the process $\phi\phi \to \chi\chi$. Since it involves $\mans$-channel contributions only, it allows for a partial-wave analysis. We will use this tool to analyse the physics content of eq.\ \eqref{eq:mixed_amp}. The differential cross section for the process $\phi\phi\to\phi\phi$ is studied in \autoref{sec:crosssection}.

\subsection{\texorpdfstring{\mans}{s}-channel scattering for distinct particles}
\label{sect:gen_s_channel}
In order to compute observables such as differential cross sections, we evaluate \eqref{eq:mixed_amp} in the centre-of-mass frame. Using the relations of \autoref{sect:conventions}, we re-express $\mant$ and $\manu$ in terms of $\mans$, the scattering angle $\theta$ and the masses $m_\phi$ and $m_\chi$.
The angular dependence of the amplitude can then readily be separated by performing a partial-wave decomposition
\be\label{partwavedef}
a_j^{\phi\chi}(\mans) \equiv \frac{1}{32 \pi} \int_{-1}^1 \mathrm{d}\cos\theta \; P_j(\cos\theta) \, \mathcal A_\mans^{\phi\chi}(\mans, \cos\theta) \, , 
\ee
where $P_j(x)$ denotes the Legendre polynomial of order $j$.

In the cases where the scattering amplitude admits resonances associated with the presence of additional massive degrees of freedom, it is also convenient to define the strength of the resonance in the spin-$j$ partial-wave amplitude via
\be\label{alphares}
\alpha_j \equiv \frac{1}{32 \pi} \int_{-1}^1 \mathrm{d}\cos\theta \; \overline{\rm Res}\left[\cA_\mans^{\phi\chi}\right] \, P_j(\cos\theta) \, ,
\ee
where
\be
\overline{\rm Res}\left[\cA_\mans^{\phi\chi}\right] = \lim_{\mans \rightarrow M^2_{\rm res}} \frac{ M^2_{\rm res} - \mans}{ M^2_{\rm res}} \, \cA_\mans^{\phi\chi} \, . 
\ee
In general, massive poles coming with a negative residue, $\alpha_j < 0$, are problematic for the unitarity of the theory \cite{Stelle:1977ry,Arici:2017whq,Alonso:2019ptb}.

It is now instructive to perform the partial-wave decomposition of the general result \eqref{eq:mixed_amp}. The angular dependence of the amplitude $\mathcal A_\mans^{\phi\chi}$ is not affected by the form factors. Thus, $\mathcal A_\mans^{\phi\chi}$ comprises only partial waves with  $j=0$ and $j=2$,
\begin{align}
 a_0^{\phi\chi}(\mans) &= \frac{1}{12} G_{RR}(\mans) \Bigg[ \left\{ (\mans+2m_\phi^2)(1+\mans \, f_{Ric\phi\phi}(\mans,m_\phi^2,m_\phi^2)) - 12 \mans \, f_{R\phi\phi}(\mans,m_\phi^2,m_\phi^2) \right\} \nonumber \\
 &\qquad\qquad\qquad \times \left\{ (\mans+2m_\chi^2)(1+\mans \, f_{Ric\chi\chi}(\mans,m_\chi^2,m_\chi^2)) - 12 \mans \, f_{R\chi\chi}(\mans,m_\chi^2,m_\chi^2) \right\} \Bigg] \, , \label{eq:partialwavesgen0} \\
 a_2^{\phi\chi}(\mans) &= -\frac{1}{60} (\mans-4m_\phi^2)(\mans-4m_\chi^2) G_{CC}(\mans) \bigg[	\left(1+\mans \, f_{Ric\phi\phi}(\mans,m_\phi^2,m_\phi^2)\right)\nonumber\\
 &\hspace{7cm}\times	\left(1+\mans \, f_{Ric\chi\chi}(\mans,m_\chi^2,m_\chi^2)\right) \bigg] \, ,
 \label{eq:partialwavesgen2}
\end{align}
whereas one has $a_j^{\phi\chi} = 0 $ for $j=1$ and $j \geq 3$. The self-interaction $\mathcal A_4^{\phi\chi}$ gives rise to partial-wave amplitudes for all even $j$. Intuitively, this corresponds to the contribution of ``ladder diagrams'' with an exchange of $n$ gravitons.

The partial-wave decomposition gives a particularly simple form for the cross section. For brevity,  we will  not discuss the contribution of the self-interaction. In the centre-of-mass frame, the cross section reads
\begin{equation}
			\left(	\frac{\mathrm{d}\sigma}{\mathrm{d}\Omega}	\right)_{\text{CM}}
	=
			\frac{1}{64\pi^2\mans}	\left|	\mathcal A	\right|^2 \, . 
\end{equation}
Inserting the partial-wave expansion of the amplitude then gives
\begin{equation}\label{eq:diffcrosssection_gen}
			\left(	\frac{\mathrm{d}\sigma_\mans^{\phi\chi}}{\mathrm{d}\Omega}	\right)_{\text{CM}}
	=
			\frac{4}{\mans}	\left|	a_0^{\phi\chi}(\mans)	P_0(\cos\theta)	+	5 a_2^{\phi\chi}(\mans)	P_2(\cos\theta)	\right|^2 \, .
\end{equation}
Due to orthogonality of the Legendre polynomials, this gives for the total cross section
\begin{equation}\label{eq:crosssection_gen}
			\sigma^{\phi\chi}_\mans
	=
			\int \mathrm{d}\Omega \left(	\frac{\mathrm{d}\sigma_\mans^{\phi\chi}}{\mathrm{d}\Omega}	\right)
	=
			\frac{16\pi}{\mans}	\left(	a_0^{\phi\chi}(\mans)^2	+	5	a_2^{\phi\chi}(\mans)^2	\right) \, ,
\end{equation}
with the partial-wave amplitudes given in eqs.\ \eqref{eq:partialwavesgen0} and \eqref{eq:partialwavesgen2}. This form of the cross section illustrates the physical number of (off-shell) graviton modes. There is a single scalar mode related to $a_0^{\phi\chi}$ and five transverse-traceless modes related to $a_2^{\phi\chi}$. This makes it clear that internal lines also depend on off-shell modes, since on-shell the graviton has only two independent polarisations.

In order to illustrate this result, we first specialise to \GR{}. In this case all gravity and gravity-matter form factors are zero.
In addition, we assume that the scalar fields are massless, \ie{}, $m_\phi=m_\chi=0$.
We also reinstate powers of $G_N$ for clarity. 
The amplitude is given by the well-known result
\be\label{grscattering}
	\cA_\mans^{\phi\chi\, \rm GR} = \, 8 \pi \, G_N \, \frac{\mant \, \manu}{\mans}
	= \, 2 \pi \, G_N \, \mans \left( 1 - \cos^2 \theta \right) \, , 
\ee
where the second equality holds in the centre-of-mass frame. The two non-vanishing amplitudes in the spin-zero and spin-two channels are
\be\label{partialwavesGR}
a_0^{\phi\chi}(\mans) = \frac{G_N}{12} \mans \, , \qquad a_2^{\phi\chi}(\mans) = - \frac{G_N}{60} \, \mans \, . 
\ee
These partial-wave amplitudes do not exhibit any poles at non-zero momentum, indicating that the scattering is solely mediated by the massless degrees of freedom described by \GR{}.
Inserting the partial-wave amplitudes into \eqref{eq:diffcrosssection_gen} and \eqref{eq:crosssection_gen} gives the differential and total cross sections
\begin{equation}
			\left( \frac{d\sigma^{\phi\chi}_\mans}{d\Omega} \right)_\text{CM}^\text{GR} = \frac{G_N^2}{16} \sin^4 \theta \, \mans \, ,
	\qquad
			\sigma^{\phi\chi}_\mans = \frac{2\pi}{15} G_N^2 \, \mans	\,.
\end{equation}
In the high-energy limit, and measured in the process energy itself, the cross section scales quadratically with the energy transfer \mans{},
\begin{equation}
 \lim_{\mans\to\infty} \left[ \mans \left( \frac{d\sigma^{\phi\chi}_\mans}{d\Omega} \right)_\text{CM}^\text{GR} \right] \propto G_N^2 \, \mans^2 \, .
\end{equation}
This points at one of the problems related to the quantisation of \GR{}: the cross section, measured in the relevant energy scale, diverges quadratically with the energy. For ``healthy'' theories it should be subject to the Froissart bound \cite{Froissart:1961ux,Froissart:2010} though, stating that $\sigma$ should not increase faster than $\log^2 \mans$. The growth of the cross section is directly linked to the fact that Newton's constant has mass-dimension $-2$. Since the scattering amplitudes are linear in $G_N$  by construction, the cross section must depend on it quadratically. The only way to balance the mass dimension introduced by $G_N$ is to have appropriate factors of the energy.

Including form factors introduces sufficient freedom to tame the growth of the amplitudes in $\mans$. Requiring that the dimensionless cross section stays finite at all energies gives conditions on the form factors. In particular, to exclude additional graviton modes we need
\begin{equation}\label{eq:noextrapoles}
 \mans f_{CC}(\mans) > -1 \, , \qquad \mans f_{RR}(\mans) > -1 \, .
\end{equation}
Requiring a bounded cross section at high energies gives a condition on the asymptotic behaviour of the propagators and form factors in the vertices. It is then convenient to introduce the asymptotic scaling laws
\be\label{scalinglaws}
\begin{aligned}
\lim_{\mans \rightarrow \infty} G_{RR}(\mans) \propto \mans^{-g_0} \, , \qquad 
& \lim_{\mans \rightarrow \infty} G_{CC}(\mans) \propto \mans^{-g_2} \, , \\
\lim_{\mans \rightarrow \infty} f_{R\phi\phi}(\mans,m_\phi^2,m_\phi^2) \propto \mans^{f_R} \, , \qquad 
& \lim_{\mans \rightarrow \infty} f_{Ric\phi\phi}(\mans,m_\phi^2,m_\phi^2) \propto \mans^{-1+f_S} \, , \\
\lim_{\mans \rightarrow \infty} f_{R\chi\chi}(\mans,m_\chi^2,m_\chi^2) \propto \mans^{c_R} \, , \qquad 
& \lim_{\mans \rightarrow \infty} f_{Ric\chi\chi}(\mans,m_\chi^2,m_\chi^2) \propto \mans^{-1+c_S} \, . \\
\end{aligned}
\ee
On this basis one can distinguish various scenarios for the large $\mans$ behaviour of the partial-wave amplitudes. The spin-two partial-wave amplitude \eqref{eq:partialwavesgen2} scales as
\be
 \lim_{\mans\to\infty} a_2^{\phi\chi}(\mans) \propto \mans^{2-g_2 + \max(0,f_S) + \max(0,c_S)} \, .
\ee
Similarly, the spin-zero partial-wave amplitude behaves as
\begin{equation}
 \lim_{\mans\to\infty} a_0^{\phi\chi}(\mans) \propto \mans^{2-g_0 + \max(0,f_S,f_R) + \max(0,c_S,c_R)} \, .
\end{equation}
Assuming that the self-interaction is sub-leading for these partial-wave amplitudes, we conclude that boundedness of the total amplitude requires that the propagators fall off at least quadratically in the squared momentum, $g_0, g_2 \ge 2$, and faster if the vertex form factors contribute. 

Based on the diagrammatic structure of the scattering amplitudes visualised in \autoref{fig:diagram_two_fields} it is clear that this analysis is actually independent of a potential momentum-dependent field redefinition of the graviton fluctuations: replacing $h \mapsto \sqrt{Z(\mans)} \,h$ with $Z(\mans)$ a positive function\footnote{Z is commonly referred to as the wave function renormalisation. In gravity it can be tensor-valued, where the defining object is actually $\left(\sqrt{Z} \right)_{\mu\nu}^{\phantom{\mu\nu}\rho\sigma}$ which maps a rank-(0,2) tensor to a rank-(0,2) tensor, including appropriate symmetries.}, each vertex receives an additional contribution $\sqrt{Z(\mans)}$ while the graviton propagator picks up an additional factor $Z(\mans)^{-1}$. Hence rescalings cancel and the amplitude remains invariant. 

\subsection{\texorpdfstring{Scattering including $\mant$- and $\manu$-channels}{Scatting including t- and u-channels}}\label{sec:crosssection}
We now consider the scattering of identical particles, $\phi\phi \rightarrow \phi\phi$. In this case the \mant{}- and \manu{}-channel also contribute to the amplitude and the partial-wave decomposition \eqref{partwavedef} is ill-defined owing to the poles in the forward and backward scattering limit.  These divergences are entirely due to the massless nature of the graviton. Since our focus is on the high-energy behaviour of the amplitudes, we will not investigate these divergences and refer to \cite{Weinberg:1965nx,Donoghue:1999qh,Akhoury:2011kq} for further discussions.

The resulting differential cross section reads 
\begin{equation}
 \left( \frac{d\sigma^{\phi\phi}}{d\Omega} \right)_\text{CM} = \frac{1}{64\pi^2 \mans} \left|\cA^{\phi\phi} \right|^2 \, ,
 \qquad \cA^{\phi\phi} \equiv \mathcal A^{\phi\phi}_\mans + \mathcal A^{\phi\phi}_\mant + \mathcal A^{\phi\phi}_\manu + \mathcal A^{\phi\phi}_4 \, .
\end{equation}

Let us first focus on the graviton-mediated part of the amplitude, comprising $\mathcal A^{\phi\phi}_\mans$, $\mathcal A^{\phi\phi}_\mant$ and $\mathcal A^{\phi\phi}_\manu$, encoded in \autoref{fig:diagrams_one_field}. Requiring that the $\mant$- and $\manu$-channel contributions do not introduce new poles for real momenta puts more stringent bounds on $f_{CC}$ and $f_{RR}$.
Since $\mant$ and $\manu$ are negative and not bounded from below, \eqref{eq:noextrapoles} needs to extend for all \emph{real} $\mans$:
\begin{equation}\label{eq:noextrapoles2}
 \mans f_{CC}(\mans) > -1 \, , \qquad \mans f_{RR}(\mans) > -1 	\qquad \text{for all } \mans\in\mathbb{R}\, .
\end{equation}

When including $\mathcal A^{\phi\phi}_\mant$ and $\mathcal A^{\phi\phi}_\manu$, we encounter a quadratic divergence in the forward scattering limit $\mans \to \infty$, $\mant$ fixed. While this will not yield a diverging cross section, it does violate causality, which requires that the amplitude grows slower than quadratically with the centre-of-mass energy \cite{Camanho:2014apa}.
At this point the form factor of the four-point interaction $f_{\phi^4}$ becomes crucial. Since the asymptotics of the graviton-mediated diagrams are fixed, the growth of the amplitude in the forward-scattering limit has to be tamed by the asymptotics of $f_{\phi^4}$:
\begin{equation}
	\lim_{\substack{\mans\to\infty\\\mant \text{ fixed}}}	\left|	\mathcal A^{\phi\phi}_\mans	+	\mathcal A^{\phi\phi}_\mant + \mathcal A^{\phi\phi}_\manu + \mathcal A^{\phi\phi}_4 \right| = o(\mans^2)\,.
\end{equation}
In \cite{Draper:2020bop}, summarised in \autoref{sect:tanh_model}, we discuss a realisation of a model where this requirement is explicitly met.

\section{The physics of graviton-mediated scattering -- examples}
\label{sect:physics_of_s_channel}
Having studied the general properties of the scattering amplitude, we will now consider specific examples corresponding to distinct quantum gravity programs. We will limit the discussion to the scattering of distinguishable particles, receiving contributions from the $\mans$-channel only, since this allows to perform a partial-wave decomposition of the amplitude. Furthermore, we will restrict ourselves to the case where the scalar fields are massless, $m_\chi = m_\phi = 0$. We will focus on scattering in the effective field theory framework (\autoref{sect:eft}), classical Stelle gravity (\autoref{sect:stelle}), infinite derivative gravity (\autoref{sect:nonlocal}), and \RG{} improvement from Asymptotic Safety (\autoref{sect:rg_improvement}).
Finally, in \autoref{sect:tanh_model} we discuss a set of form factors that lead to scattering amplitudes which are scale-free at trans-Planckian energies without introducing any poles for real squared momenta and, as a result, satisfy all constraints regarding unitarity and causality. 
Our results are summarised in \autoref{tab:examples}.

\begin{table}
	\centering
	\begin{tabular}{p{.25\textwidth}p{.15\textwidth}p{.23\textwidth}p{.25\textwidth}}
		\hline
			theory	&	gravity \newline form factors 	&	pole structure	&	\UV{} behaviour	\\
		\hline\hline
			effective field theory	&	$\log$	&	pole at cutoff scale	&	n/a	\\
		\hline
				Stelle gravity	&	$\text{const}$	&	massive spin-two \newline ghost d.o.f.	&	const \newline scale-free	\\
		\hline
			infinite derivative \newline gravity	&	$\exp$	&	essential \newline singularity at $\infty$	&	$\exp$ falloff (\mans\ ch.),\newline $\exp$-divergent ($\mant/\manu$ ch.)\\
		\hline
			Asymptotic Safety:\newline \RG{} improvement	& const	&		massive spin-two \newline ghost d.o.f.	&	const \newline scale-free	\\
		\hline
			Asymptotic Safety:\newline form factor model	&	$\tanh$	&	infinite tower at \newline imaginary \newline squared momentum	& const \newline	scale-free	\\
		\hline\hline
	\end{tabular}
	\caption{Characteristic features of the scattering amplitudes obtained from the quantum gravity models discussed in \autoref{sect:physics_of_s_channel}. The column ``gravity form factors'' gives the functional form of $f_{RR}$ and $f_{CC}$ in the corresponding theory. The columns ``pole structure'' and ``\UV{} behaviour'' describe the pole structure of the graviton propagator for complexified momenta and the asymptotics of the $\mans$-channel amplitude for large centre-of-mass energy, respectively.}\label{tab:examples}
\end{table}

\subsection{Effective field theory and IR properties}
\label{sect:eft}
The differential cross section calculated in the previous section readily incorporates results obtained by effective field theory \cite{Donoghue:1994dn}. In this approach one takes the viewpoint that a quantum theory is valid up to a \UV{} cutoff scale $\Lambda_{\text{UV}}$.
Starting with a bare action $S$, one may then compute the cross section by including perturbative (one-loop) corrections to the tree-level Feynman diagrams \cite{Donoghue:1993eb,Donoghue:1994dn,Barvinsky:1994cg,Khriplovich:2002bt,BjerrumBohr:2002kt,Codello:2015oqa,Ohta:2020bsc}, also see \cite{Burgess:2003jk} for a pedagogical introduction. These corrections comprise non-analytic contributions proportional to $\log \mathfrak{s}$ and $\mathfrak{s}^{-1/2}$ \cite{Donoghue:1993eb,Donoghue:1994dn}.

Alternatively, the one-loop form factors can be calculated directly from the one-loop effective action
\begin{equation}
	\label{eq:1loopea}
		\Gamma^{\rm 1-loop}
	=
		S
	-	\frac{1}{2}	\tr \log	\frac{S^{(2)}+\mathcal{R}_{\Lambda_{\text{UV}}}}{S^{(2)}}
	\,,
\end{equation}
where $S^{(2)}$ is the second variation with respect to the fluctuation fields. Furthermore, we introduced the cutoff operator $\mathcal{R}_{\Lambda_{\text{UV}}}$ to regularise the trace. For $S$ being the Einstein-Hilbert action supplemented by minimally coupled scalar fields, the universal part of the gravitational form factors is 
\begin{equation}\label{eq:1loopgrav}
		\Gamma^{\rm non-local} =	\frac{1}{32\pi^2}	\int \text{d}^4x \sqrt{-g}	\left[
				c_1	R \, \log\left(	\frac{\Delta}{\Lambda_{\text{UV}}^2}	\right)	R
			+	c_2 \, C_{\mu\nu\rho\sigma}	\log\left(	\frac{\Delta}{\Lambda_{\text{UV}}^2}	\right)	C^{\mu\nu\rho\sigma}
			\right]
	\,.
\end{equation}
For pure gravity one has $c_1 = 1/4$ and $c_2 = 7/20$ \cite{Satz:2010uu,Percacci:2017fkn} while the Einstein-Hilbert action supplemented by $N_s$ minimally coupled, massless scalar fields yields $c_1 = 1/4 + N_s/72$ and $c_2 = 7/20 + N_s/120$. 
Matter form factors have been given in \cite{Codello:2015oqa,Ohta:2020bsc}.

By construction, effective field theory becomes invalid at energies $\mathfrak{s}\gtrsim \Lambda_{\text{UV}}^2$. Thus, these form factors only capture the low-energy behaviour. In order to access the high-energy limit of the form factors, different approaches have to be used.

While this work is mostly concerned with \UV{} properties of amplitudes and cross sections, let us briefly comment on \IR{} effects. Generically non-localities appear in the form factors if a theory contains massless modes, \eg{}, the logarithms above. Both fundamental and phenomenological aspects of more general non-local terms have been discussed in the literature \cite{Codello:2015mba, Codello:2015pga, Belgacem:2017cqo, Belgacem:2020pdz, deBrito:2020wmp}. Some non-local terms may have interesting consequences for the universe at large scales \cite{Wetterich:1997bz, Maggiore:2013mea, Maggiore:2014sia}, potentially providing an explanation of dark energy. A recent reconstruction of the form factor $f_{RR}$ from non-perturbative Monte Carlo simulations \cite{Knorr:2018kog} lends support for this idea also from first principles.

\subsection{Stelle gravity}
\label{sect:stelle}

\begin{figure}[b!]
	\centering
		\includegraphics[width=.6\textwidth]{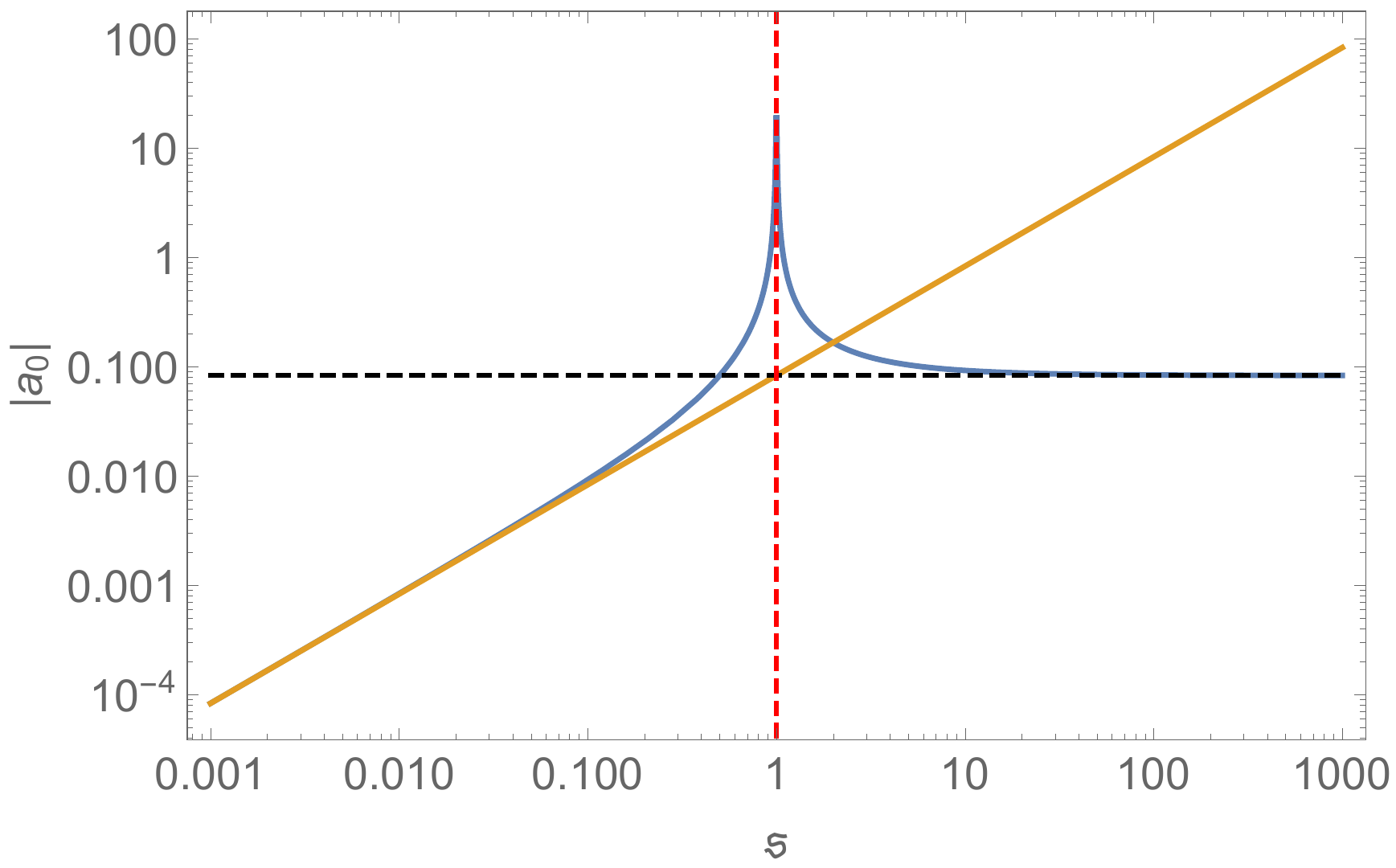}
	\caption{
	Illustration of the partial-wave amplitude $a_0$ associated with the process $\phi\phi\to \chi\chi$ in classical Stelle gravity (eqs.\ \eqref{eq:stellepartialwaves}) and from the \RG{} improvement (eqs.\ \eqref{partialwavesimp}) for $c_R = 1$ and $g_* = 1$, respectively. The dashed red line highlights the position of the pole $\mans=c_R$, triggering the transition to the regime where the amplitude is scale-free. The asymptotic value of the amplitude is visualised by the dashed horizontal line. 	
	 The amplitude for \GR{} is depicted by the orange line, indicating that the results agree for $\mans \lesssim 1/10$.
	The amplitude $a_2$ for $c_C=1$ is obtained by the rescaling $a_2 = -a_0/5$.
	}
	\label{fig:stelle_cross_section}
\end{figure}

As a second application of our general results we consider classical Stelle gravity \cite{Stelle:1976gc,Stelle:1977ry}. This theory possesses several attractive features, such as perturbative renormalisability. Stelle gravity is obtained from the action \eqref{eq:Gammagrav} by setting the form factors $f_{RR}$ and $f_{CC}$ to a constant:
\begin{equation}
f_{RR}	=	-	c_R
\,,	\qquad
f_{CC}	=	-	c_C
\,.
\end{equation}
For the gravity-matter interaction we implement minimal coupling. The free parameters $c_R, c_C > 0$ control the scale where the modifications owed to the higher-derivative terms set in.
Inserting the form factors into the partial-wave expressions \eqref{eq:partialwavesgen0} and \eqref{eq:partialwavesgen2} we obtain
\begin{equation}\label{eq:stellepartialwaves}
a_0^{\text{Stelle}} = \frac{1}{12} \mans^2 \left[ \frac{1}{\mans} - \frac{1}{\mans-c_R^{-1} } \right] \, , \\
\qquad	
a_2^{\text{Stelle}} = -\frac{1}{60} \mans^2 \left[ \frac{1}{\mans} - \frac{1}{\mans-c_C^{-1} } \right] \, .
\end{equation}
As expected, the presence of the quadratic curvature terms changes the behaviour of the amplitudes at high energies, rendering them finite.
Since quadratic gravity contains four derivatives acting on the metric, it is expected that this theory gives rise to additional degrees of freedom. In the partial-wave decomposition, this is reflected by poles appearing at finite energy. Indeed, the expressions \eqref{eq:stellepartialwaves} diverge at $\mans = c_R^{-1}$ and $\mans= c_C^{-1}$, respectively. This is illustrated in \autoref{fig:stelle_cross_section}.
While the additional pole in the scalar sector corresponds to a healthy degree of freedom, the pole in $G_{CC}$ is associated with a negative-energy state. This is known to violate unitarity \cite{Stelle:1977ry}. Formally, this can be shown by calculating the residues associated with the massive poles \cite{Arici:2017whq,Alonso:2019ptb}
\begin{equation}
\alpha_0^{\text{Stelle}} = \frac{1}{12 c_R} > 0 \, , 
\qquad
\alpha_2^{\text{Stelle}} = -\frac{1}{60 c_C} < 0 \, .
\end{equation}
Since $\alpha_2^{\text{Stelle}}$ is negative, the massive spin-two pole corresponds to an Ostrogradski ghost signalling that the theory is problematic (also see \cite{Anselmi:2017ygm, Anselmi:2018tmf, Anselmi:2019xac, Donoghue:2019ecz} for recent reinterpretations of this degree of freedom as virtual particle leading to the violation of microcausality).

\subsection{Infinite Derivative Gravity}
\label{sect:nonlocal}

By construction, infinite derivative gravity (\IDG) includes form factors already at the level of the bare action. A typical choice for the functions determining the flat-space propagators is \cite{Talaganis:2014ida, Ghoshal:2017egr}
\begin{equation}\label{IDG-ff}
 f_{RR} = \frac{e^{c_R \Delta}-1}{\Delta} \, , \qquad f_{CC} = \frac{e^{c_C \Delta}-1}{\Delta} \, , \qquad f_{\phi\phi} = e^{c_S (\Delta - m_\phi^2)} \, (\Delta-m_\phi^2) \, .
\end{equation}
The parameters $c_R, c_C$, and $c_S$ set the scale where the form factors give relevant contributions. Typically, these are identified, $c_R = c_C = c_S$, and expressed in terms of the non-locality scale $M^2 \equiv (c_R G_N)^{-1}$.
The exponentials are designed to regulate the high-energy behaviour of loop diagrams, yielding  a theory which is renormalisable and asymptotically free at high energy \cite{Biswas:2011ar}. Studies at the level of infinite derivative scalar theories \cite{Talaganis:2014ida, Talaganis:2016ovm} furthermore indicate that taming the growth of all scattering amplitudes requires introducing form factors for the matter vertices. For the gravity-matter sector given in \eqref{eq:gamma-matter} the corresponding terms have not been worked out. As these are not critical for analysing gravity-mediated $\mans$-channel scattering, we will set the corresponding form factors to zero and work with minimally coupled scalar fields.

\begin{figure}
	\centering
  				\includegraphics[width=0.6\textwidth]{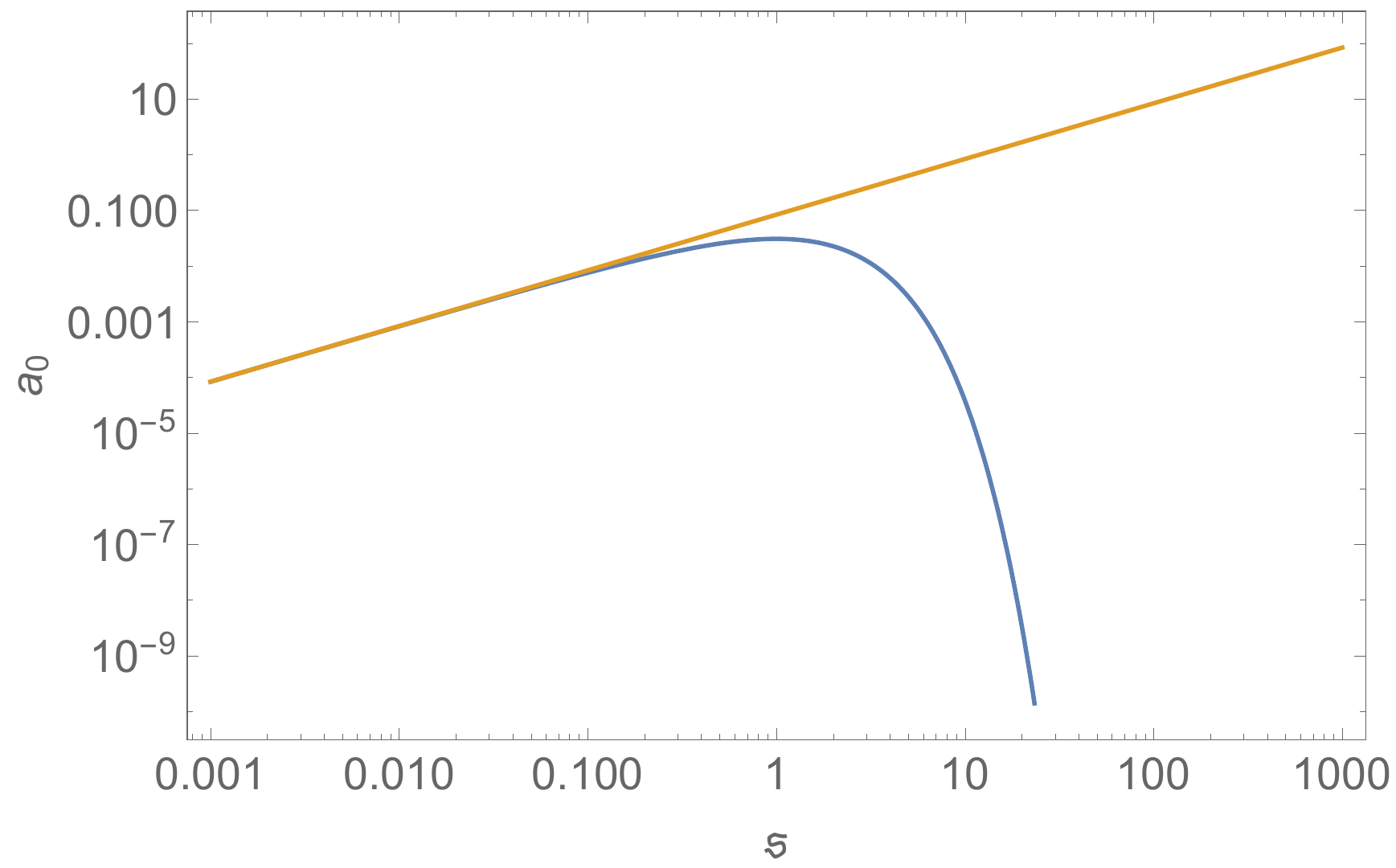}
\caption{\label{fig:idgamplitude} Illustration of the partial-wave amplitude $a_0$, eq.\ \eqref{eq:idgpwamplitude}, resulting from infinite derivative gravity with $c_R = 1$ (blue line). The corresponding result from \GR{} is given by the orange line for comparison. For energies below the non-locality scale, $ \mans \lesssim 1$, the two amplitudes coincide while for $\mans \gtrsim 1$ the infinite derivative gravity amplitude decreases exponentially. The amplitude $a_2$ for $c_C=1$ is obtained by the rescaling $a_2 = -a_0/5$.
}
\end{figure}

The form factors \eqref{IDG-ff} appear at the level of the bare action and are thus subject to renormalisation. Assuming that \eqref{IDG-ff} carries over to the quantum effective action, the resulting partial-wave amplitudes are
\begin{equation}\label{eq:idgpwamplitude}
 a_0^{\rm IDG} = \frac{1}{12} \mans \, e^{- \, c_R\, \mans} \, , \qquad a_2^{\rm IDG} = -\frac{1}{60} \mans \, e^{-\, c_C \,\mans} \, .
\end{equation}
Comparing to \eqref{partialwavesGR}, we observe that these are the partial-wave amplitudes found in \GR{} multiplied by an exponential factor. In the \mans-channel, this results in an amplitude that decreases exponentially as $\mans\to\infty$, as is pictured in \autoref{fig:idgamplitude}.\footnote{When considering the scattering of identical particles $\phi\phi\to\phi\phi$, crossing symmetry entails that the $\mant$- and $\manu$-channels contain exponentials whose arguments are given by $\mant$ and $\manu$. Since $\mant$ and $\manu$ are negative, this results in a cross section that diverges exponentially. This can be amended by choosing exponentials with quadratic arguments, \ie{}, Gaussian form factors. For a discussion of even and odd polynomials in the exponential, see also \cite{Buoninfante:2018mre}. } Notably, the partial waves are regular on the positive real axis, $\mans > 0$, indicating that the theory does not give rise to additional massive degrees of freedom.

\subsection{Renormalisation Group improvements from Asymptotic Safety}
\label{sect:rg_improvement}

We will now turn our attention to form factors from non-perturbative renormalisation group (\RG) flows. First, we will look at the process of ``\RG{} improvement'' \cite{Dittrich:1985yb}. The key idea stems from particle physics, and we will quickly illustrate it for the well-known Uehling potential. One starts with a physical quantity that one wants to \RG{}-improve, \ie{}, where one wants to include the leading order quantum corrections, for example the Coulomb potential,
\begin{equation}
 V(r) = -\frac{e^2}{4\pi r} \, .
\end{equation}
The \RG{} improvement step then consists in replacing the charge $e$ by the running gauge coupling while identifying the \RG{} scale with the relevant physical scale, here the radius $r$. In this way, at one-loop order,
\begin{equation}
 V(r) \mapsto -\frac{e^2(r_0^{-1})}{4\pi r} \left[ 1 + \frac{e^2(r_0^{-1})}{6\pi^2} \log\left( \frac{r_0}{r}\right) + \mathcal O(e^4) \right] \, ,
\end{equation}
where $r_0$ is an \IR{} reference scale. This gives the correct one-loop Uehling potential \cite{Uehling:1935uj}, which can also be obtained by more conventional perturbative methods \cite{Dittrich:1985yb}.
In a similar spirit, \RG{} improvements in gravity have been used by various groups to study potential phenomenological consequences of Asymptotic Safety in cosmology \cite{Bonanno:2001xi,Reuter:2005kb,Hindmarsh:2012rc,Bonanno:2017pkg,Bonanno:2018gck} and black hole physics \cite{Bonanno:2000ep,Cai:2010zh,Falls:2012nd,Koch:2014cqa,Platania:2019kyx,Bonanno:2019ilz}, also see \cite{Donoghue:2015nba,Donoghue:2019clr,Bonanno:2020bil} for discussions on the limitations of this strategy.

Typically, the asymptotic safety program \cite{Percacci:2017fkn,Reuter:2019byg} computes the dependence of couplings on a coarse-graining scale $k$. The effective coupling $G(k)$ includes quantum fluctuations with momenta $p^2 \ge k^2$. The Newton constant $G_N$, appearing in the effective action \eqref{eq:Gammagrav}, is obtained in the limit $k \rightarrow 0$ where \emph{all quantum fluctuations are included}. In the simplest case the $k$-dependence of $G(k)$ can be computed from projecting the Wetterich equation \cite{Wetterich:1992yh,Morris:1993qb,Reuter:1996cp} onto the (Euclidean) Einstein-Hilbert action \cite{Reuter:1996cp,Reuter:2001ag}. Neglecting the effect of a cosmological constant, this leads to a flow equation for the dimensionless Newton's coupling $g_k \equiv k^2 G(k)$, 
\be
k \partial_k g_k = (2 + \eta_N) g_k \, , \qquad \eta_N = \frac{B_1 \, g}{1-B_2 \, g}\,.
\ee
Here $B_1$ and $B_2$ are numerical coefficients which, in general, depend on the regularisation procedure. Their precise values are not important in the present discussion and it suffices to note that they are numbers of $\mathcal{O}(1)$. Neglecting the terms proportional to $B_2$ in the denominator, the flow equation admits the analytic solution \cite{Bonanno:2000ep}
\be\label{RGimp1}
G(k^2) = \frac{G_N}{1+\omega \, G_N \, k^2} \, , \qquad \omega = - \frac{1}{2} \, B_1 \, . 
\ee
The flow of $G(k^2)$ then interpolates between the fixed point value $\lim_{k \rightarrow \infty} g_k = g_* \equiv \omega^{-1}$ and $\lim_{k \rightarrow 0} G(k^2) = G_N$. The function $G(k^2)$ is regular for all values $k^2 \ge 0$.

When applying the \RG{} improvement procedure to the classical amplitude \eqref{grscattering}, one replaces $G_N \rightarrow G(k^2)$ and subsequently identifies the coarse-graining scale with a physical momentum scale. For \mans{}-channel scattering the natural choice is the momentum transfer \mans{} carried by the graviton which cuts off the propagator in the infrared. Exploiting that the flat Euclidean background admits a standard Wick rotation suggests the cutoff-identification
\be\label{cutoffid}
k^2 \mapsto - \mans \, . 
\ee
As a result one then arrives at the \RG{}-improved amplitude
\be\label{Ampimp}
\begin{split}
	\cA_\mans^{\rm RG-imp} = & \, \frac{8 \pi }{1 - \omega \, \mans} \, \frac{\mant \, \manu}{\mans} = 8 \pi \, \mant \manu \, \left[ \frac{1}{\mans} - \frac{1}{\mans - g_* } \right] \, . \\
\end{split}
\ee
The regularity of $G(k^2)$ for $k^2 \ge 0$ then ensures that the \RG{} improvement does not introduce unphysical poles situated at negative values of $\mans$.

The amplitude \eqref{Ampimp} gives two non-vanishing partial-wave amplitudes for spin zero and spin two:
\be\label{partialwavesimp}
a_0^{\text{RG-imp}} = \frac{1}{12} \mans^2 \left[ \frac{1}{\mans} - \frac{1}{\mans - g_* } \right] 
 \, , \quad a_2^{\text{RG-imp}} = - \frac{1}{60} \, \mans^2 \, \left[ \frac{1}{\mans} - \frac{1}{\mans - g_*} \right] . 
\ee
The form factors ``created'' by the \RG\ improvement can then be determined by comparing the partial-wave amplitudes to the general expressions \eqref{eq:partialwavesgen0} and \eqref{eq:partialwavesgen2},
\be\label{ffRGimp}
f_{RR} = f_{CC} = - \omega \, . 
\ee
The comparison to \eqref{eq:stellepartialwaves} indicates that the \RG{} improvement actually gives rise to a specific realisation of Stelle gravity in which $c_R = c_C = g_*^{-1}$ are fixed by the position of the \RG{} fixed point. Hence, the partial-wave amplitudes again take the form illustrated in \autoref{fig:stelle_cross_section}. Furthermore, all remarks made in the context of Stelle gravity carry over to this case as well. In particular, one again has massive poles in $a_0^{\text{RG-imp}}$ and $a_2^{\text{RG-imp}}$ which ensure that the amplitudes approach constant values as $\mans \rightarrow \infty$.
Thus we conclude that the \RG{} improvement based on the simple formula \eqref{RGimp1} actually manages to tame the growth of the amplitude at large $\mans$, incorporating a key feature of Asymptotic Safety. At the same time the ansatz does this in the simplest possible way, introducing one additional massive degree of freedom in the spin-zero and spin-two channel. We expect that this feature is actually signalling a deficit in the \RG{} improvement procedure which will not hold in a first-principle computation.

At this stage the following cautious remark related to applying the \RG{} improvement procedure to more general scattering processes is in order. Generically,  form factors come with multiple arguments, indicating that they depend on several, independent scales. Thus there is no natural identification of the \RG{} scale with a physical momentum scale, and the map between the several-parameter form factors and the $k$-dependent coupling cannot be one-to-one. As a consequence, it is expected that the procedure breaks down when we consider multi-particle scatterings and higher-order vertices. This is in line with arguments put forward in \cite{Donoghue:2015nba,Donoghue:2019clr}. In fact, we can see this limitation of the \RG{} improvement already in the case discussed here: the \RG{} improvement assigns the same functional dependence to both the spin-two and spin-zero sector of the off-shell graviton, which generically are not the same in a more general calculation with individual form factors. It is also clear from a fundamental point of view that whereas in electrodynamics, \RG{} improvement at the level of the action can be straightforwardly carried out,
\begin{equation}
 \frac{1}{e^2} F_{\mu\nu} F^{\mu\nu} \mapsto F_{\mu\nu} \frac{1}{e^2(\Delta)} F^{\mu\nu} \, ,
\end{equation}
a similar strategy seems not feasible in the case of a running Newton's, or cosmological constant \cite{Hamber:2013rb}.

We close this subsection stressing that there has been a significant effort in determining the momentum dependence of propagators and vertices from first-principle computations based on solutions of the Wetterich equation for \emph{Euclidean signature} \cite{Wetterich:1992yh,Morris:1993qb,Reuter:1996cp}. While these computations have not reached the level where the gauge-invariant scattering amplitudes investigated in this work can be studied in detail, it is nevertheless worthwhile to summarise the status of these efforts in order to identify the missing links.

Currently, there are two paths towards studying momentum-dependent correlation functions followed in the literature, so-called background calculations and fluctuation calculations. On the background side, the first study of form factors with the \FRG{} re-derived the Polyakov effective action in two dimensions \cite{Codello:2010mj}. The first non-perturbative calculation of a form factor has been performed in \cite{Bosma:2019aiu} within a conformally reduced setting, whereas \cite{Franchino-Vinas:2018gzr} defined a running Newton's constant from the form factor related to the linear term, that is in terms of total derivatives. Form factors have also been considered for gravity coupled to matter, see e.g. \cite{Codello:2015oqa} for a comprehensive overview of different matter fields, and \cite{Knorr:2019atm} for a calculation of the propagator of a scalar field coupled to gravity. On the side of fluctuation calculations, most works employ a flat background. As shown in eqs.\ \eqref{eq:grav_propagator} and \eqref{propparameterization}, the form factors of the quadratic curvature terms can be mapped one-to-one to the graviton propagator. The fully momentum-dependent propagator of the graviton was first calculated in \cite{Christiansen:2014raa}. Follow-up works also investigated partial momentum dependences of the three- \cite{Christiansen:2015rva} and four-point function \cite{Denz:2016qks, Eichhorn:2018akn}, as well as gravity-matter systems \cite{Christiansen:2017cxa}. Notably, \cite{Christiansen:2017bsy, Burger:2019upn} performed a calculation of momentum-dependent propagators on a background with constant curvature.

Most of the calculations have in common that they only resolve the propagators. The discussion around eq.\ \eqref{scalinglaws} shows that those alone are not sufficient to make statements about scattering amplitudes. One can always perform a momentum-dependent field redefinition to bring the propagator into standard form, at the cost of introducing extra momentum dependence in the vertices. This observation has been extensively used in the fluctuation calculations, and the general observation was that the remaining momentum dependence of the three- and four-graviton vertex seems to be weak. A map of the correlators discussed in these works onto their diffeomorphism invariant form factor expressions is currently missing and constitutes an area for future research. 

\subsection{Form factors realising Lorentzian Asymptotic Safety}
\label{sect:tanh_model}
The discussion of the previous subsections revealed that in many approaches to quantum gravity, the resulting scattering amplitudes exhibit additional massive degrees of freedom at Planckian energy. This raises the question whether the quantum effective action can accommodate amplitudes which are bounded everywhere and scale-free at high energy. In \cite{Draper:2020bop} this question was answered in the affirmative. The physics ingredient underlying such models is an infinite tower of massless (Lee-Wick type) poles located on the imaginary axis of the complex $\mans$-plane which exhibits a Regge-type scaling behaviour asymptotically. A concrete realisation of this mechanism is provided by the gravitational form factors
\begin{equation}\label{infiniteres}
\begin{aligned}
f_{RR}(\Delta)
=
c_{R}	\, \tanh \left(	c_{R} 	\, \Delta	\right)
\,,
\qquad
f_{CC}(\Delta)
=
c_{C} 	\, \tanh	\left(	c_{C}	\, \Delta	\right) \, .
\end{aligned}
\end{equation}
The construction is accompanied by a four-point vertex associated with the scalar self-interactions, eq.\ \eqref{eq:4mixed}, whose contribution to the amplitude is given by
\be
\mathcal A_4^{\phi\chi} = 4\pi G_C(\mans) (\mant^2+\manu^2) f^{\text{int}}(\mans^2+\mant^2+\manu^2) \, ,
\ee
with the interpolation function being	
	\be\label{intfct}
	f^{\rm int}(x) = \frac{c_t \, x \tanh[c_t \, x]}{1 + c_t \, x \tanh[c_t \, x]} \, . 
	\ee
The choice \eqref{intfct} ensures that $\mathcal A_4^{\phi\chi}$ is invariant under crossing symmetry and that the self-interaction does not contribute at low energy. The two numerical parameters $c_{R}, c_{C} \ge 0$ control the position of the imaginary poles in the graviton propagator while $c_t$ sets the scale where the self-interactions start contributing to the amplitude.
The forward-scattering limit of the amplitude $\mathcal A^{\phi\phi}$ is made finite by the self-interaction.
In the present discussion, this contribution is not needed and thus was not included in the analysis.

The key feature of \eqref{infiniteres} is that it leads to a modified gravitational propagator which falls of as $\mans^{-2}$ while being well-defined on the entire real axis apart from a first order pole at $\mans = 0$. Hence, the model has the same degrees of freedom as \GR{}.
\begin{figure}
	\centering
\includegraphics[width=0.6\textwidth]{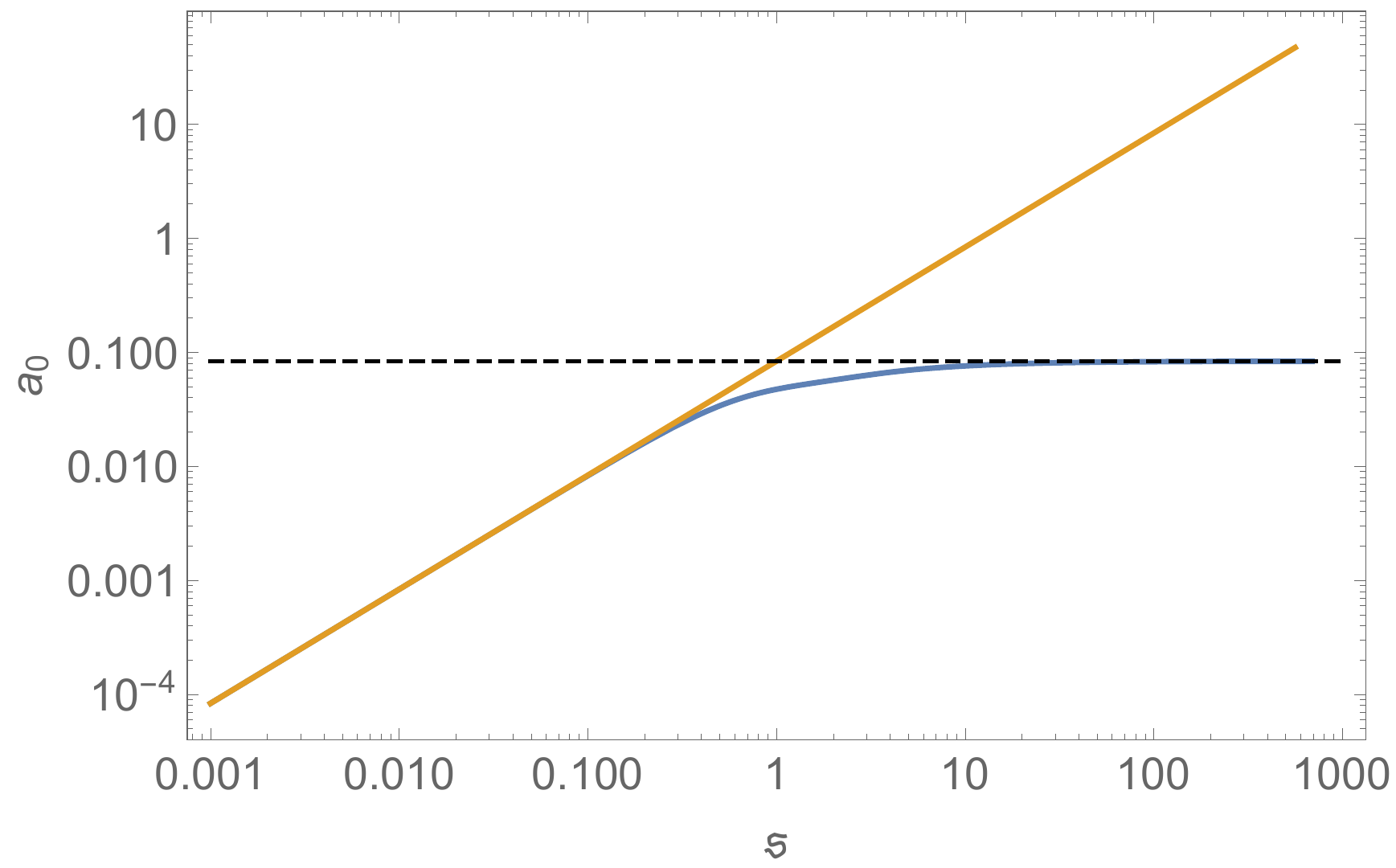}
\caption{Illustration of the partial-wave amplitude $a_0$ resulting from \eqref{partialwavesirm} with $c_R = 1$ (blue line). The corresponding result from \GR{} is given by the orange line for comparison. The asymptotic value of the partial-wave amplitude is displayed by the horizontal dashed line. The amplitude $a_2$ for $c_C=1$ is obtained by the rescaling $a_2 = -a_0/5$.
}
	\label{fig:IRM_cross_section}
\end{figure}
The partial-wave amplitudes arising from \eqref{infiniteres} are
\be\label{partialwavesirm}
\begin{split}
a_0
= \frac{1}{12} \, \frac{ \mans}{1 + c_R \mans \tanh(c_R \mans)} \, , 
\qquad a_2
= - \frac{1}{60} \, \frac{ \mans}{1 + c_C \mans \tanh(c_C \mans)} .
\end{split} 
\ee
Their shape depends on $c_C $ and $c_R $ only and it is illustrated in \autoref{fig:IRM_cross_section}.
The form factors in the denominators tame the growth of the amplitude for large $\mans$ so that the partial-wave amplitudes approach a constant value for $\mans \to \infty$. 
This behaviour is triggered by the poles located at imaginary squared momentum $\mans$ and does not require the introduction of massive gravitational modes.

\section{Summary and Discussion}\label{sect:summary}
In this paper, we have used the form factor parameterisation of the quantum effective action $\Gamma$ to construct  the most general amplitude for a two-scalar-to-two-scalar process mediated by gravitons in a Minkowski background. Our result  covers the most general momentum dependence for the graviton and scalar propagators as well as  the scalar-scalar-graviton vertices and scalar self-interactions. The explicit expressions for the amplitudes describing the scattering of distinguishable particle species  and a single particle species are given in eqs.\ \eqref{eq:mixed_amp} and \eqref{eq:Amp_phi_gen}, respectively. They exhibit the following  properties:
\begin{enumerate}
	\item[1)]	The amplitudes are independent of the gauge parameters introduced in eq.\ \eqref{eq:gaugefixing}, demonstrating the gauge invariance of our result. Gauge independence is recovered in a two-step process: first, the three-point functions project out the unphysical scalar and vector modes from the propagator. Secondly, any gauge dependence in the physical scalar and tensor modes are projected away by evaluating the amplitude on-shell.
	\item[2)] The scattering amplitude is invariant under (momentum-dependent) field redefinitions of internal lines. Any redefinition of the graviton fluctuations leads to a momentum-dependent inverse wave function renormalisation in the propagator. At the level of  the amplitude this factor is cancelled by identical contributions coming from the vertices.
	\item[3)] The amplitudes are derived in Lorentzian signature. Hence, no prescription for a Wick rotation is required. 
\end{enumerate}
In particular, properties 1) and 2) ensure that we are dealing with valid physical observables. This allows us to implement fundamental requirements related to unitarity and causality directly at the level of these observables which then translate into constraints on the quantum effective action.

The general expression for the partial-wave amplitudes associated with gravity-mediated $\phi\phi \to \chi\chi$-scattering are given in eqs.\ \eqref{eq:partialwavesgen0} and \eqref{eq:partialwavesgen2}. Inserting specific choices for the form factors then allowed us to contrast the amplitudes resulting from different quantum gravity models. The resulting insights can be summarised as follows:

\begin{enumerate}
	\item The form factor formulation naturally includes the one-loop effective action arising from the effective field theory framework of general relativity \cite{Donoghue:1993eb,Donoghue:1994dn}, including the universal logarithmic correction terms \cite{Satz:2010uu,Codello:2015oqa,Ohta:2020bsc}. These form factors capture the universal \IR{} behaviour, valid at energy scales well below the \UV{}-cutoff scale.

	\item We gave a form factor description for classical Stelle gravity \cite{Stelle:1977ry}. The quadratic curvature terms introduce additional massive poles (ghosts) in the partial-wave spectrum. At energies above these poles, the amplitudes become scale-free. The pole in the $j=2$ amplitude comes with a negative residue, illustrating a violation of unitarity. In this context, it is a rather remarkable observation that starting from the amplitudes computed in \GR{} and performing the simplest \RG{} improvement motivated by Asymptotic Safety recovers the form factors of Stelle gravity and fixes the free parameters in terms of the  renormalisation group fixed point underlying Asymptotic Safety. At the same time, we have shown that the \RG{} improvement scheme is not able to capture the full information encoded in the form factors: since \RG{} improvement is limited to a single scale identification, the two different off-shell modes of gravity as well as the form factors related to vertices with multiple arguments cannot be described in a one-to-one way. This also resolves the puzzle raised in \cite{Donoghue:2015nba,Donoghue:2019clr}.

	\item Infinite derivative gravity \cite{Talaganis:2014ida,Ghoshal:2017egr} introduces form factors in such a way that the propagators obtained in \GR{} are dressed by exponential prefactors without introducing additional massive poles at finite momentum. As a result, the $\mans$-channel amplitudes fall of exponentially for centre-of-mass energy above the non-locality scale. The $\mant$-channel amplitude obtained by crossing symmetry grows exponentially, so that it is unclear whether this class of models is free from \UV{} divergences. 
	
	\item Finally, we reviewed the model introduced in \cite{Draper:2020bop}, exhibiting scattering amplitudes which are bounded everywhere and scale-free in the \UV{} without introducing new massive degrees of freedom or ghosts. The taming of the \UV{} behaviour is generated by the collective interplay of an infinite tower of massless (Lee-Wick type) poles in the propagator and suitable form factors in the matter sector. This ensures that the construction obeys unitarity and causality conditions.

\end{enumerate}

In conclusion, we have introduced a new perspective on gravity-mediated particle scattering within the framework of relativistic quantum field theory. Our results provide 
guidelines for the construction of well-behaved scattering amplitudes satisfying unitarity and causality constraints. We expect that these are useful for top-down constructions of quantum gravity starting from a microscopic description as well as for quantum gravity model-building. In particular, they serve as a \emph{proof of principle} that the Lorentzian quantum effective action is able to accommodate scattering amplitudes which obey the requirements of Asymptotic Safety \cite{Weinberg:1976xy,Weinberg:1980gg}. This construction \emph{does not include massive higher-spin resonances}. This is in contrast to stringy \UV{} completions such as the Veneziano and Virasoro-Shapiro amplitudes \cite{Veneziano:1968yb, Virasoro:1969me, Shapiro:1970gy}, which provide a \UV{} completion of general relativity by introducing an infinite tower of higher-spin degrees of freedom \cite{Green:1987sp}.

\appendix
\section{Explicit expressions for propagators and vertices}
\label{App.A}
In this appendix, we collect explicit expressions for the graviton propagator and scalar-scalar-graviton vertex, computed from the quantum effective action \eqref{eq:Gans1}.

\subsection{Graviton propagator}\label{app:gravitonpropagator}
We first list the expression for the graviton propagator, which we compute from \eqref{eq:Gammagrav} supplemented by the gauge fixing action \eqref{eq:gaugefixing}. This yields
\begin{equation}\label{eq:grav_propagator}
\begin{aligned}
 \frac{1}{\mathbf i} G^{\mu\nu}_{\phantom{\mu\nu}\rho\sigma}(p) &= \delta^{(\mu}_\rho \delta^{\nu)}_\sigma G_{\mathbbm 1}(p^2) + \frac{1}{4} \eta^{\mu\nu} \eta_{\rho\sigma} \, G_\text{Tr}(p^2) + \frac{p^\mu p^\nu p_\rho p_\sigma}{p^4} \, G_{p^4}(p^2) \\
 & \quad + \frac{1}{2} \frac{ \eta^{\mu\nu} p_\rho p_\sigma + p^\mu p^\nu \eta_{\rho\sigma} }{p^2} G_{gpp}(p^2) + \frac{p^{(\mu}_{\phantom{)}} \delta^{\nu)}_{(\rho} p^{\phantom{)}}_{\sigma)}}{p^2} G_{pgp}(p^2) \, ,
\end{aligned}
\end{equation}
with the scalar propagator functions
\begin{align}
 G_{XX}(p^2) &= \frac{1}{p^2 \left( 1+ p^2 f_{XX}(p^2) \right)} \, , \nonumber\\
 G_{\mathbbm 1}(p^2) &= 32\pi G_{CC}(p^2) \, , \nonumber\\
 G_\text{Tr}(p^2) &= -\frac{128 \pi }{3} G_{CC}(p^2) - \frac{64\pi }{3} G_{RR}(p^2) \, , \nonumber\\
 G_{p^4}(p^2) &= -\frac{64\pi }{p^2} \frac{\alpha (\beta-1)(\beta-5)}{(\beta-3)^2} + \frac{64\pi }{3} G_{CC}(p^2) - \frac{256\pi }{3} \frac{\beta^2}{(\beta-3)^2} G_{RR}(p^2) \, , \nonumber\\
 G_{gpp}(p^2) &= \frac{64\pi }{3} G_{CC}(p^2) + \frac{128\pi }{3} \frac{\beta}{\beta-3} G_{RR}(p^2) \, , \nonumber\\
 G_{pgp}(p^2) &= \frac{64\pi }{p^2} \alpha - 64\pi G_{CC}(p^2) \, . \label{propparameterization}
\end{align}
We can see that the identity and trace component are gauge invariant, and that there is a singularity for the gauge choice $\beta=3$. This choice is related to an incomplete gauge fixing -- in that case, the gauge fixing operator is actually a projector \cite{Gies:2015tca}.

\subsection{Gravity-matter vertices}\label{app:vertices}
Here we list the explicit formulas of the gravity-matter vertex and the matter four-point vertex originating from \eqref{eq:gamma-matter}. First, we state the gravity-matter three-point vertex:
\begin{align}
 \Gamma_\phi^{(h\phi\phi)\,\mu\nu} (p_h, p_{\phi_1}, p_{\phi_2}) &= \frac{1}{4} \eta^{\mu\nu} \bigg( f_{\phi\phi}(p_{\phi_1}^2) + f_{\phi\phi}(p_{\phi_2}^2) \bigg) \notag \\
 & - \frac{1}{2} \frac{f_{\phi\phi}(p_{\phi_1}^2) - f_{\phi\phi}(p_{\phi_2}^2)}{p_{\phi_1}^2 - p_{\phi_2}^2} \Bigg[ p_{\phi_1}^\mu p_{\phi_1}^\nu + p_{\phi_2}^\mu p_{\phi_2}^\nu - \frac{1}{2} \eta^{\mu\nu} p_h \cdot (p_{\phi_1} + p_{\phi_2}) \notag \\
 & \hspace{4.5cm} + \frac{1}{2} \left( p_h^\mu (p_{\phi_1}^\nu + p_{\phi_2}^\nu) + p_h^\nu (p_{\phi_1}^\mu + p_{\phi_2}^\mu) \right) \Bigg] \notag \\
 &+ \bigg( f_{R\phi\phi}(p_h^2,p_{\phi_1}^2,p_{\phi_2}^2) + f_{R\phi\phi}(p_h^2,p_{\phi_2}^2,p_{\phi_1}^2) \bigg) \bigg( p_h^2 \eta^{\mu\nu} - p_h^\mu p_h^{\nu\phantom{\mu}} \hspace{-0.1cm} \bigg) \notag \\
 &- \frac{1}{2} f_{Ric\phi\phi}(p_h^2,p_{\phi_1}^2,p_{\phi_2}^2) \bigg[ \frac{1}{2} \left( p_h^2 + p_{\phi_1}^2 - p_{\phi_2}^2 \right) \left( p_h^\mu p_{\phi_1}^\nu+ p_h^\nu p_{\phi_1}^\mu \right) \notag \\
 & \hspace{4.5cm} + \frac{1}{4} \left( p_h^2 + p_{\phi_1}^2 - p_{\phi_2}^2 \right)^2 \eta^{\mu\nu} + p_h^2 p_{\phi_1}^\mu p_{\phi_1}^\nu \bigg] \notag \\
 &- \frac{1}{2} f_{Ric\phi\phi}(p_h^2,p_{\phi_2}^2,p_{\phi_1}^2) \bigg[ \frac{1}{2} \left( p_h^2 + p_{\phi_2}^2 - p_{\phi_1}^2 \right) \left( p_h^\mu p_{\phi_2}^\nu+ p_h^\nu p_{\phi_2}^\mu \right) \notag \\
 & \hspace{4.5cm} + \frac{1}{4} \left( p_h^2 + p_{\phi_2}^2 - p_{\phi_1}^2 \right)^2 \eta^{\mu\nu} + p_h^2 p_{\phi_2}^\mu p_{\phi_2}^\nu \bigg] \, . \label{eq:vertex}
\end{align}
In this expression, $p_h$ corresponds to the momentum of the graviton and $p_{\phi_{1,2}}$ are the momenta of the scalars. The finite difference in the second line is characteristic for the variation of form factors and takes the place of the naively expected derivative \cite{Knorr:2019atm}. We will assume that the scalar kinetic form factor $f_{\phi\phi}$ is differentiable so that the vertex is finite everywhere.

Finally, the four-scalar vertices read
\begin{equation}\begin{aligned}
		\Gamma_\phi^{(\phi^2\chi^2)}(p_1,p_2,p_3,p_4) &= f_{\phi^2\chi^2}(p_1\cdot p_2, p_1 \cdot p_3, p_1 \cdot p_4, p_2 \cdot p_3, p_2 \cdot p_4, p_3 \cdot p_4) \, ,
 \\
		\Gamma_\phi^{(\phi^4)}(p_1,p_2,p_3,p_4) &= f_{\phi^4}(p_1\cdot p_2, p_1 \cdot p_3, p_1 \cdot p_4, p_2 \cdot p_3, p_2 \cdot p_4, p_3 \cdot p_4) \, .
\end{aligned}\end{equation}
Let us recall that we assumed that $f_{\phi^2\chi^2}$ is symmetric under interchanging $p_1 \leftrightarrow p_2$ and $p_3\leftrightarrow p_4$, whereas $f_{\phi^4}$ is assumed to be symmetric in all its arguments.

\section{Conventions}\label{sect:conventions}

In this appendix we summarise our conventions. To make contact with standard \QFT{} literature, we employ a mostly-minus convention for the signature of the metric. In this way, the Minkowski metric reads
\begin{equation}
 \eta_{\mu\nu} = \text{diag}(1,-1,-1,-1) \, .
\end{equation}
With this, the on-shell condition for a free scalar field of mass $m$ reads
\begin{equation}
 p^2 = p_0^2 - \mathbf p^2 = m^2 \, .
\end{equation}
When no indices are used, vectors in standard font correspond to four-vectors, whereas bold-face vectors correspond to three-vectors. In this way, the four-momentum of a free scalar field reads
\begin{equation}
 p_\mu = \left( \sqrt{m^2 + \mathbf p^2}, \mathbf p \right) \, .
\end{equation}
At any vertex, all momenta are defined as ingoing,
\begin{equation}
 \sum_i p_i = 0 \, .
\end{equation}
Finally, we define the centre-of-mass frame. For two different scalar fields $\phi$ and $\chi$ with masses $m_\phi$ and $m_\chi$ in an \mans{}-channel scattering, we choose a frame in which
\begin{align}
 p_{1\mu} &= \left(\sqrt{m_\phi^2+\mathbf{p}^2}, \mathbf p\right) \, , \;
 &  p_{2\mu} & =  \left(\sqrt{m_\phi^2+\mathbf{p}^2}, -\mathbf p\right) \, , \\
 p_{3\mu} &= \left(-\sqrt{m_\chi^2+\mathbf{q}^2}, \mathbf q\right) \, , \; 
 & p_{4\mu} & =  \left(-\sqrt{m_\chi^2+\mathbf{q}^2}, -\mathbf q\right) \, .
\end{align}
By four-momentum conservation, we find that the magnitudes of the three-momenta are related by
\begin{equation}
 \mathbf q^2 = m_\phi^2 - m_\chi^2 + \mathbf p^2 \, .
\end{equation}
This must be positive and gives restrictions on the kinematically allowed minimal momentum transfer. The three-momentum $\mathbf q$ furthermore defines the scattering angle $\theta$ with respect to $\mathbf p$ via
\begin{equation}
 \mathbf p \cdot \mathbf q = \sqrt{\mathbf p^2 \mathbf q^2} \cos \theta \, .
\end{equation}
Finally, we can define the Mandelstam variables in this frame:
\be\label{eq:mansdef}
\begin{split}
 \mans &= (p_1+p_2)^2 =4(m_\phi^2 + \mathbf p^2) \, , \\
 \mant &= (p_1+p_3)^2 = -\mathbf p^2 - \mathbf q^2 - 2 \sqrt{\mathbf p^2 \mathbf q^2} \cos\theta \, , \\
 \manu &= (p_1+p_4)^2 = -\mathbf p^2 - \mathbf q^2 + 2 \sqrt{\mathbf p^2 \mathbf q^2} \cos\theta \, .
\end{split}
\ee
We can also express the squared three-momenta in terms of \mans{},
\begin{equation}
 \mathbf p^2 = \frac{1}{4} \mans - m_\phi^2 \, , \quad \mathbf q^2 = \frac{1}{4} \mans - m_\chi^2 \, ,
\end{equation}
and insert this into the expressions for \mant{} and \manu{},
\begin{align}
 \mant &= - \left( \frac{\mans}{2} - m_\chi^2 - m_\phi^2 + \frac{1}{2} \sqrt{\left(\mans - 4m_\phi^2\right)\left(\mans - 4m_\chi^2\right)} \cos\theta \right) \, , \\
 \manu &= - \left( \frac{\mans}{2} - m_\chi^2 - m_\phi^2 - \frac{1}{2} \sqrt{\left(\mans - 4m_\phi^2\right)\left(\mans - 4m_\chi^2\right)} \cos\theta \right) \, .
\end{align}
In this way we can write the scattering amplitudes as a function of \mans{} and the scattering angle $\theta$. We can also check the well-known identity
\begin{equation}\label{eq:msrel}
 \mans + \mant + \manu = p_1^2 + p_2^2 + p_3^2 + p_4^2 \equiv 2 \left( m_\phi^2 + m_\chi^2 \right) \, ,
\end{equation}
indicating the Mandelstam variables satisfy a linear constraint equation.

\acknowledgments{We would like to thank Wim Beenakker, Anupam Mazumdar, Alessia Platania, Les\l{}aw Rachwa\l{}, Martin Reuter, and Melissa van Beekveld for interesting discussions. B.\ K.\ acknowledges support by Perimeter Institute for Theoretical Physics. Research at Perimeter Institute is supported in part by the Government of Canada through the Department of Innovation, Science and Economic Development Canada and by the Province of Ontario through the Ministry of Colleges and Universities. F.\ S.\ acknowledges financial support from the Netherlands Organisation for Scientific Research (NWO) within the Foundation for Fundamental Research on Matter (FOM) grant 13VP12.
}

\bibliographystyle{JHEP}
\bibliography{general_bib}

\end{document}